\documentclass{article}
\usepackage[a4paper]{geometry}
\usepackage[utf8]{inputenc}
\usepackage[T1]{fontenc}     
\usepackage[english]{babel}	
\usepackage{amsmath,amssymb, amsthm} 
\usepackage[font=scriptsize]{caption}
\usepackage{epsfig}
\usepackage{mathrsfs}
\usepackage{cite}
\usepackage{xcolor}
\usepackage{hyperref}

\def\N{\mathbb{N}}
\def\R{\mathbb{R}}

\newcommand{\norm}[1]{\left\lVert#1\right\rVert}      

\theoremstyle{plain}
\newtheorem{theorem}{Theorem}

\theoremstyle{definition}
\newtheorem{definition}{Definition}

\theoremstyle{remark}

\def\A{\mathcal{A}}
\def\Q{\mathcal{Q}}
\DeclareMathOperator{\Id}{Id}

\begin{document}
\title{\bf Braids with the symmetries of Platonic polyhedra in the Coulomb (N+1)-body problem}
\author{Marco Fenucci\footnote{Dipartimento di Matematica,
    Universit\`a di Pisa, Largo B. Pontecorvo 5, 56127 Pisa, Italy}
\and 
        Àngel Jorba
        \footnote{Departament de Matemàtiques i Informàtica, 
Barcelona Graduate School of Mathematics (BGSMath), 
Universitat de Barcelona (UB), 
Gran Via de les Corts Catalanes 585, 
08007 Barcelona, 
Spain}}
\date{4 November 2019}

\maketitle
\begin{abstract}
We take into account the Coulomb $(N+1)$-body problem 
with $N=12,24,60$. One of the particles has positive charge $Q>0$,
and the remaining $N$ have all the same negative charge $q<0$. 
These particles move under the Coulomb force, and the positive charge
is assumed to be at rest at the center of mass. 
Imposing a symmetry constraint, given by the symmetry group of the Platonic polyhedra, 
we were able to compute periodic orbits, using a shooting method and continuation with 
respect to the value $Q$ of the positive charge. 

In the setting of the classical $N$-body problem, the existence of 
such orbits is proved with Calculus of Variation techniques, by minimizing the action functional. Here this 
approach does not seem to work, and numerical computations show that
the orbits we compute are not minimizers of the action.



\end{abstract}

\section{Introduction}
The classical Newtonian $N$-body problem, for its history and its challenges, is a
fundamental problem in mathematical physics. During the centuries, many mathematicians
faced it, starting with Newton, who formulated the law of universal gravitation and
solved the problem of two bodies, passing from Lagrange, who was the first to discover the triangular solutions of the three-body
problem, through Poincar\'e which discovered the chaotic nature of the problem. 
In more recent years, the discovery of the Figure Eight solution of the three-body problem
\cite{moore93} started a research on particular periodic orbits of the $N$-body problem,
that now are known with the name of \textit{choreographies}. Numerical evidence of the
existence of such orbits are given in \cite{simo2000, simo2002, simo:newfamilies}, 
and also rigorous computer assisted methods to study their existence and their dynamical properties have
been used \cite{kapela02, kapela:eight, kapela:KAM, FG18}. The first rigorous proof of the
existence of the Figure Eight orbit was given in \cite{CM2000}, and it uses the method of
minimization of the action over a particular set of periodic loops. Since then, this
technique has been used to prove the existence of other several periodic motions, see 
\cite{CM2000, hiphop, FT2004, terracini, terrvent07, fgn10, FG18, marchal01, ch02, chen03,
Chen2008} and references therein. The idea of minimizing the action goes
back to Poincar\'e \cite{poincare1896}.

Connection between the macroscopic scale, i.e. celestial mechanics, and the microscopic
scale, i.e. the atomic mechanics, is given by the Coulomb force, which governs the interaction
between charged particles. The Coulomb potential depends on $1/r$, as the Newtonian
potential, with the difference that it depends on the charges and not on the masses. 
However, charges can be also negative, making this force both attractive and repulsive.  
The most famous connection is given by the Rutherford model of the atom
\cite{rutherford11}, which represents the atom as a miniature Solar System. Since
the introduction of quantum mechanics, this model was deprecated. However, despite quantum
mechanics provides a more accurate description of nature, classical methods are still useful in
studying atomic dynamics \cite{uzer91}. 
In more recent years, the Coulomb $N$-body problem, i.e. the problem of $N$ charged particles
which interact through the Coulomb force, was taken into account, and
periodic orbits in the Coulomb three-body problem were found, see for example \cite{perez1996, santander1997,
sindikEtAt2018}, trying to reproduce some features of the known orbits in the
classical Newtonian three-body problem. 
On the other hand, some special symmetric solutions were found numerically in
\cite{davies83}, for small values of $N$.
The use of the Coulomb force as the only
interacting force is motivated by the fact that the gravitational interaction is 
negligible when the electrostatic force is introduced in the system.
Moreover, the Coulomb force by itself amounts to a non-relativistic approximation, which is reliable only
when the velocities are small compared to the speed of light \cite{Landau1980Classical}.

In this paper we consider the Coulomb $(N+1)$-body problem, composed by a positive
charged particle and other $N$ negative equally charged particles. We search for
periodic motions sharing the symmetry of Platonic polyhedra, that is
\textit{Tetrahedron, Cube, Octahedron, Dodecahedron} and \textit{Icosahedron}, 
hence $N$ can be either $12,24$ or $60$. For the Newtonian $N$-body problem, a list of orbits with
this symmetry was found in \cite{FG18}. Here we have been able to
compute a similar set of periodic orbits, see the webpage
\cite{MFwebCoulomb} for
animations of these solutions. Our numerical computations also show that
these orbits are unstable.
Moreover, since the approach used in
\cite{fgn10, FG18} for the proof of the existence was the minimization of the action, here we
investigated also if this method could still work for the Coulomb $(N+1)$-body problem.
However, we show numerically that the orbits we compute are not minimizers of the action, not
even locally.

The paper is organized as follows. Section~\ref{s:model} 
introduces the model used. In Section~\ref{s:numcomp} we describe the numerical methods
used for the computation and in Section~\ref{s:results} we present the results obtained.

\section{The Coulomb $(N+1)$-body problem}
\label{s:model}
%
We take into account a system composed by $N+1$ charged particles, one of which has
positive charge and the rest have equal negative charges. Despite the Rutherford model
\cite{rutherford11} turned out to be not valid to represent the physical nature of the
atom, for the sake of simplicity we will use terms as \textit{electrons} and
\textit{nucleus} in the following.

We denote with $q<0, m>0$ the charge and the mass of the electron respectively, with $Q>0$ 
the charge of the nucleus, with $u_i \in \R^3, i=1,\ldots,N$ the position of the
$i$-th particle and with $u_0 \in \R^3$ the position of the nucleus. 
The mass of the nucleus is very high compared with the mass of the electrons, so we assume
that the nucleus stays fixed at $u_0\in\R^3$.
The particles move under the Coulomb force, and the system of equations that determine the motion is given by 
\begin{equation}
   m \ddot{u}_i = \kappa q \bigg[ \sum_{\substack{j=1 \\ j\ne i}}^N
      q\frac{u_i-u_j}{|u_i-u_j|^3}
   +  \ Q\frac{u_i-u_0}{|u_i-u_0|^3} \bigg], \quad i=1,\dots,N,
   \label{eq:coulombODE}
\end{equation}
where $\kappa > 0$ is the Coulomb constant.

We choose a reference frame centered at the center of mass, hence 
from now on we assume that $u_0\equiv 0$.
The system \eqref{eq:coulombODE} is Lagrangian, and the Lagrangian $L$ is given by
\[
   L = K + U,
\]
where 
\begin{equation}
   K = \frac{1}{2}\sum_{i=1}^N m|\dot{u}_i|^2,
   \label{eq:kin}
\end{equation}
is the kinetic energy and
\begin{equation}
   U = -\kappa \sum_{1\leq i < j \leq N} \frac{q^2}{|u_i-u_j|} - \kappa\sum_{i=1}^N
   \frac{qQ}{|u_i|},
   \label{eq:pot}
\end{equation}
is the Coulomb potential. To simplify the computations, we can choose the units of
charge, mass and distance so that
\begin{itemize}
   \item[-] the charge of the electron is unitary, hence $q=-1$,
   \item[-] the mass of the electron is unitary, hence $m=1$,
   \item[-] the Coulomb constant is unitary, hence $\kappa = 1$.
\end{itemize}
Solutions to the equations \eqref{eq:coulombODE} can be found also as stationary points of
the Lagrangian action functional, defined as
\begin{equation}
   \mathcal{A}(u) = \int_{a}^{b}L(u,\dot{u}) \,dt.
   \label{eq:actFun}
\end{equation}
The functional \eqref{eq:actFun} is defined over a set of curves $\mathcal{K} \subseteq
H^1([a,b], \R^{3N})$, which has to be specified, depending on the problem that one wants to study. 

\subsection{Symmetry of the Platonic polyhedra and topological constraints}\label{sec:spp}
We want to compute periodic orbits of the system \eqref{eq:coulombODE}, imposing both
symmetrical and topological constraints.
As done in \cite{fgn10, FG18}, we take into account a Platonic polyhedron
(i.e. \textit{Tetrahedron, Cube, Octahedron, Dodecahedron} and \textit{Icosahedron}) and 
we denote with $\mathcal{R}$ its rotation group. We consider a system composed by
$N = |\mathcal{R}|$ electrons (hence $N$ can be either $12,24$ or $60$) and, identifying $\{ 1,\cdots,N \}$ with
the elements of $\mathcal{R}$, we label the positions of the particles with the rotations of the group. 
We search for periodic orbits of period $T>0$ such that
\begin{itemize}
   \item[(a)] the motion $u_R$, $R \in \mathcal{R} \setminus \{ I \}$ is recovered by
      \[
         u_R(t) = R u_I(t), \quad t \in \R;
      \]
   \item[(b)] the trajectory of $u_I$, that we call \emph{generating particle},
      belongs to a given non-trivial free-homotopy class of $\R^3 \setminus \Gamma$, where
      \[
         \Gamma = \bigcup_{R \in \mathcal{R} \setminus \{ I\}} r(R), 
      \]
      and $r(R)$ is the axis of the rotation $R$. Note that, by condition (a),
      $\Gamma \setminus \{0\}$ corresponds to the set of the partial collisions between the
      electrons;
   \item[(c)] there exist $R \in \mathcal{R}$ and $M>0$ such that
      \[
         u_I(t+T/M) = R u_I(t), \quad t \in \R.
      \]
\end{itemize}
In \cite{fgn10, FG18} periodic orbits of the $N$-body problem with equal masses,
satisfying these symmetries and topological constraints, were found as minimizers of the Lagrangian action functional, 
using Calculus of Variations techniques in order to prove their existence. 

In our case, taking into account the symmetry (a), the action functional writes as 
\begin{equation}
   \mathcal{A}(u) = N \int_{0}^{T}\bigg( \frac{1}{2}|\dot{u}_I|^2 - \frac{1}{2}
   \sum_{R \in \mathcal{R} \setminus\{ I\}} \frac{1}{|(R-I)u_I|} + \frac{Q}{|u_I|} \bigg)
   \,
   dt,
   \label{eq:actionSym}
\end{equation}
and it is defined on the set 
\begin{equation}
   \mathcal{K} = \{ u \in H_T^1(\R, \R^{3N}) : \text{ (a), (b) and (c) hold} \},
   \label{eq:loopSpace}
\end{equation}
where $H^1_T(\R,\R^{3N})$ is the space of $T$-periodic loops of $H^1(\R, \R^{3N})$.
Since a term with a negative sign appears in the Lagrangian, it is not clear whether this
functional is coercive or not, and the search for periodic orbits using the usual 
minimization of the action technique does not apply so easily. For this reason we investigate the existence 
of these periodic orbits with a preliminary numerical study.

Note that \eqref{eq:actionSym} depends only on the path of the generating
particle $u_I$. This means that we can reduce the searching for periodic orbits of the
full system of charges to the searching of periodic orbits of the generating particle
$u_I$, whose dynamics is defined by the Lagrangian
\begin{equation}
   L = \frac{1}{2}|\dot{u}_I|^2 - \frac{1}{2}
   \sum_{R \in \mathcal{R} \setminus\{ I\}} \frac{1}{|(R-I)u_I|} + \frac{Q}{|u_I|}.
   \label{eq:redLag}
\end{equation}
The Euler-Lagrange equations of \eqref{eq:redLag}, written as first order system, are
\begin{equation}
   \begin{cases}
      \displaystyle \dot{u}_I = v_I, \\[2ex]
      \displaystyle \dot{v}_I = \sum_{R \in \mathcal{R}\setminus\{ I \}} \frac{(I-R)u_I}{|(R-I)u_I|^3} -
      \frac{Q u_I}{|u_I|^3}.
   \end{cases}
   \label{eq:redEqMotion}
\end{equation}
This system has the advantage that the dimension is much smaller than the dimension
of the system of equations \eqref{eq:coulombODE}, $6$ compared to $6N$ . Moreover, if the periodic orbit of the
generating particle $u_I$ is unstable in the system \eqref{eq:redEqMotion}, then the
complete orbit with $N$ electrons is unstable in the system \eqref{eq:coulombODE}. 
On the other hand, the stability in the reduced system leads only to the 
stability with respect to symmetric perturbations of the complete orbit. To study entirely the
stability, we need to solve equations \eqref{eq:coulombODE}, together with its
variational equation, in order to compute the full $6N \times 6N$ monodromy matrix. Since we do not
expect to find many stable orbits for the reduced system, the study of the stability is
divided in two steps: at the first step we check whether the generating particle is stable
in the reduced system or not and, if it is stable, we proceed in the computation of the full
monodromy matrix and get an estimation the Floquet multipliers.
\section{Numerical methods}
\label{s:numcomp}
In this section we describe the methods used to compute periodic orbits of
\eqref{eq:coulombODE} with the constraints described in Section \ref{s:model}.  
The main method is a variant of the well-known shooting method, but here problems arise
when we search for a good starting guess. In \cite{FG18} the starting guess was computed using a
gradient descent method, applied to the discretized version of the action functional
of the $N$-body problem. In our case we cannot use the same method, because the lack of 
coercivity of the action \eqref{eq:actionSym} leads to a failure of the gradient
method, that we experienced in our numerical experiments. A good
starting guess is found applying a continuation method to a
modification of the initial problem.

\subsection{Shooting method}
\label{ss:shooting}
In order to compute the orbits, we use a shooting
method in the phase space of the generating particle. The goal is to solve the boundary
value problem
\begin{equation}
   \begin{cases}
      \dot{x} = f(x), \\
      x(T/M) = Sx(0),
   \end{cases}
   \label{eq:bvp}
\end{equation}
where $f$ is the vector field in \eqref{eq:redEqMotion}, $x \in \R^6$, $S$ is the matrix
\[
   S = 
   \begin{pmatrix}
      R & 0 \\
      0 & R
   \end{pmatrix},
\] 
and $R \in \mathcal{R}$, $M > 0$ are given by condition (c) in
Section~\ref{sec:spp}. Fixed $n$ values
$0 = \tau_0 < \tau_1 < \cdots < \tau_{n}=T/M$, we define
the function $G:\R^{6n} \to \R^{6n}$ as
\begin{equation}
   \begin{cases}
      G_i = \phi^{\tau_i - \tau_{i-1}}(x_{i-1}) - x_i, & i=1,\dots,n-1 \\
      G_n = \phi^{\tau_n-\tau_{n-1}}(x_{n-1}) - Sx_0.
   \end{cases}
   \label{eq:shootingEq}
\end{equation}
If we have a $T$-periodic solution $x(t)$ satisfying \eqref{eq:bvp},
the function $G$ evaluated at
\[
X = (x(\tau_0), \dots, x(\tau_{n-1}))
\]
vanishes. Zeros of the function \eqref{eq:shootingEq} are thus computed with a Newton
method. The Jacobian matrix of $G$ is
\begin{equation}
   \frac{\partial G}{\partial X } = 
   \begin{bmatrix}
      M_1 & -\Id & & \\
               & M_2 & \ddots & & \\ 
               & & \ddots & -\Id \\
               -S & & & M_n
   \end{bmatrix},
   \label{eq:jacG}
\end{equation}
where 
\[
   M_i = \frac{\partial}{\partial x} \phi^{\tau_{i}-\tau_{i-1}}(x_i).
\]
If $X'$ denotes the new value of $X$ at some iteration of the Newton method and
$\Delta X = X'-X$, at each step we solve the linear system
\begin{equation}
   \frac{\partial G}{\partial X}(X) \Delta X = -G(X).
   \label{eq:linSys}
\end{equation}
However, the Jacobian matrix is singular at zeros of $G$, since we are free to
choose the initial point along the periodic orbit. This degeneracy
can be avoided as in \cite{barrio:periodic}, by adding a transversality condition on the first shooting point
\begin{equation}
   f(x_0) \cdot \Delta x_0 = 0,
   \label{eq:ortDisp}
\end{equation}
to the system \eqref{eq:linSys}, where $x_0, \Delta x_0$ are the first components
of $X, \Delta X$ respectively.  The system of equations \eqref{eq:linSys}, \eqref{eq:ortDisp} has
$6n+1$ equations and $6n$ unknowns, and we can solve it through the SVD
decomposition, thus obtaining the value of $\Delta X$.

Moreover, in order to make the method more stable, we choose to use a damped Newton method,
i.e. the new value $X'$ at the generic iteration is obtained as
\begin{equation}
   X' = X + \gamma \Delta X.
   \label{eq:dampedNewton}
\end{equation}
The damping parameter $\gamma$ is adaptive, since it is computed as
\[
  \gamma =\displaystyle\frac{\gamma_{\text{min}}}
          {\max\big(\gamma_{\text{min}},\, |\Delta X|_{\infty}\big)},
\]
where $|\Delta X|_\infty = \max_i |\Delta X_i|$. In our software we set
$\gamma_{\text{min}}=\frac{1}{10}$.

\subsection{Continuation method}
\label{ss:continuation}
Suppose now that the vector field in \eqref{eq:bvp} also depends on a real parameter,
say $\lambda$, hence $f=f(x,\lambda)$. In this manner, the function $G$ in
\eqref{eq:shootingEq} also depends on $\lambda$, hence $G=G(X,\lambda)$. 
%
%
As we will use the continuation method for different purposes, we explain it using a
generic continuation parameter $\lambda \in \R$.
Given a couple $(X_i, \lambda_i)^T$ such that $G(X_i, \lambda_i)=0$, we want to continue
this solution with respect to the varying parameter $\lambda$, in order to find a curve of
solutions parametrized with $\lambda$.
%
To this end, we add to the system \eqref{eq:shootingEq} an equation to displace the entire couple
$(X,\lambda)^T$. The system we solve is therefore
\begin{equation}
   \begin{cases}
      \displaystyle G(X,\lambda) = 0, \\[2ex]
      \displaystyle \bigg|
      \begin{pmatrix}
         X_i\\ \lambda_i
      \end{pmatrix} -
      \begin{pmatrix}
         X \\ \lambda
      \end{pmatrix}
      \bigg|^2 - \delta^2 = 0,
   \end{cases}
   \label{eq:continuationEq}
\end{equation}
where, as usual, $|\,\cdot\,|$ denotes the Euclidean norm and $\delta > 0
$ is a positive parameter determining the displacement along the curve
of solutions.  The solution $(X_{i+1}, \lambda_{i+1})^T$ of system
\eqref{eq:continuationEq} is computed using again a Newton method,
solving at each step a linear system given by the matrix
\begin{equation}
   \begin{bmatrix}
      \displaystyle \frac{\partial G}{\partial X} & \displaystyle \frac{\partial
      G}{\partial \lambda} \\[2ex]
      2(X-X_i) & 2(\lambda-\lambda_i)
   \end{bmatrix}.
   \label{eq:newtonContinuation}
\end{equation}
The initial guess
can be
constructed starting from the known solution $(X_i, \lambda_i)^T$, and
taking a tangent displacement along the curve of solutions. However,
we have simply constructed the initial guess by approximating this
tangent line using two previous different solutions, say $(X_i,
\lambda_i)^T$ and $(X_{i-1},\lambda_{i-1})^T$, as
\[
\begin{pmatrix}
   \hat{X} \\
   \hat{\lambda}
\end{pmatrix} 
=
\begin{pmatrix}
   X_i \\ \lambda_i
\end{pmatrix}
+
\gamma 
\begin{pmatrix}
   X_{i} - X_{i-1} \\
   \lambda_{i} - \lambda_{i-1}
\end{pmatrix}, 
\qquad 
\gamma = \frac{\delta}{|(X_i,\lambda_i)^T - (X_{i-1}, \lambda_{i-1})^T|}.
\]
As we are searching for periodic solutions, for the reasons explained above, we
have added the transversality condition \eqref{eq:ortDisp}. The final system we
have to solve at each step is non-squared, and we use again the
SVD decomposition to solve it.

Note also that, since $G$ is defined through the flow of an ordinary differential
equation and we need to compute the derivatives of $G$ with respect to the parameter
$\lambda$, we have to compute the derivatives of the flow with respect to $\lambda$. To do
this, the system of equations that we have to solve numerically is
\begin{equation}
   \begin{cases}
      \displaystyle \dot{x} = f(x,\lambda), \\[2ex]
      \displaystyle \dot{A} = \frac{\partial f(x, \lambda)}{\partial x} A, \\[2ex]
      \displaystyle \dot{w} = \frac{\partial f(x,\lambda)}{\partial x} w +
      \frac{\partial f(x,\lambda)}{\partial \lambda},
   \end{cases}
   \label{eq:vEq}
\end{equation}
where $x,w \in \R^6, A \in \R^{6 \times 6}$ and $\lambda \in \R$. Indeed, the second
equation gives the derivatives of the flow $\phi^t$ with respect to the initial condition
$x$, while the third equation gives the derivatives of the flow $\phi^t$ with respect to the
parameter $\lambda$.

\subsection{Computing periodic orbits}
\label{ss:startingGuess}
As it has been mentioned before, one of the difficulties to carry out
these computations is to find a good initial guess for the Newton
method to converge. To deal with this, we have used two different
continuation schemes.

From the physical intuition, if the central charge is zero, the electrons only repel each other,
and we do not expect to find any periodic solutions. For continuity reasons, if the 
central charge is too small compared to the number of electrons, a periodic orbit could
still not exist. On the contrary, if the positive charge is large enough, the contribution
of the electrons in the vector field \eqref{eq:coulombODE} is small, 
compared to the term given by the positive charge. 
Indeed, rescaling the loops as $u_i(t) = Q^{1/3}v_i(t), \
i=1,\dots,N$ in \eqref{eq:coulombODE}, we obtain a differential equation for 
$v_i$, which writes as
\[
   \ddot{v}_i = \mu \sum_{\substack{j=1 \\ j\neq i} }^N
   \frac{v_i-v_j}{|v_i-v_j|^3}- \frac{v_i}{|v_i|^3}, \quad i=1,\cdots,N,
\]
where $\mu = Q^{-1/3}$. Note that, when the positive charge is ideally
``infinite'', the interactions between the different electrons disappear, and the
differential equation that determines the motion of $v_i$ is the equation of a Kepler
problem.
Intuitively, when the central charge is finite but very large, the solution is close to a circular piecewise
loop, composed by Keplerian arcs, joined at points on the collision lines. 
Therefore, physical intuition suggests that periodic solutions are more likely to exist when the central
charge is large enough.
For all these reasons it is easier to find periodic orbits for high values of the central
charge and to continue them to lower values of $Q$. 

To find periodic orbits for large values of $Q$ we use the following strategy.
First we choose a closed curve 
\begin{gather*}
   \varphi:[0,T] \to \R^6, \qquad \varphi(t)= \big( u(t), \dot{u}(t) \big)^T,
\end{gather*}
such that the image of the spatial component $u([0,T])$ belongs to the chosen
free-homotopy class of $\R^3 \setminus \Gamma$. Moreover, we
can take this spatial component to be on a sphere, since
we expect that the final orbit does not have large changes in the radial component.
Of course this curve will not solve equation \eqref{eq:bvp}, 
but we can perturb the system and define a new differential equation for which
$\varphi$ is a solution. Indeed, if we define
\begin{equation}
   \dot{x} = f(x) - \varepsilon \psi(t), 
   \label{eq:perturbedEq}
\end{equation}
where
\[
 \psi(t) =  \dot{\varphi}(t)-f(\varphi(t)),
\]
then $\varphi(t)$ is a solution of \eqref{eq:perturbedEq} for $\varepsilon = 1$.
To find the periodic orbit for $\varepsilon=0$, we consider $\varepsilon$ as a parameter
and we use the continuation method. 
In our computations, we decided to stop the continuation when we
reach a value of $\varepsilon < 10^{-2}$: this was usually enough to have an initial guess 
for the shooting method to converge for $\varepsilon=0$ and compute the periodic orbits.

Summarizing, the approach used to compute the orbits is the divided in three steps. 
\begin{itemize}
   \item[Step 1:] We generate a starting guess for a high value of the central charge
      $Q$, with the method described above. In our computations, we
      decided to choose a value near $2N$, i.e. two times the number of electrons.
   \item[Step 2:] We compute the solution using the shooting method described in
      Section \ref{ss:shooting}, with the same value of $Q$ used to
      produce the initial guess, and using continuation
      w.r.t. $\varepsilon$ as explained above, until we reach $\varepsilon=0$.
      Using this last solution as starting guess, we compute a second solution for the value of
      the central charge equal to $Q-1$: this is needed to start the continuation w.r.t.
      the value of the central charge.
   \item[Step 3:] Using the two solutions computed at the Step 2, we start the
      continuation method described in Section \ref{ss:continuation}, in order to find
      solutions for smaller values of $Q$.
\end{itemize}

\section{Results of the computations}
\label{s:results}

\begin{figure}[!ht]
   \centerline{
      \includegraphics[scale=0.68]{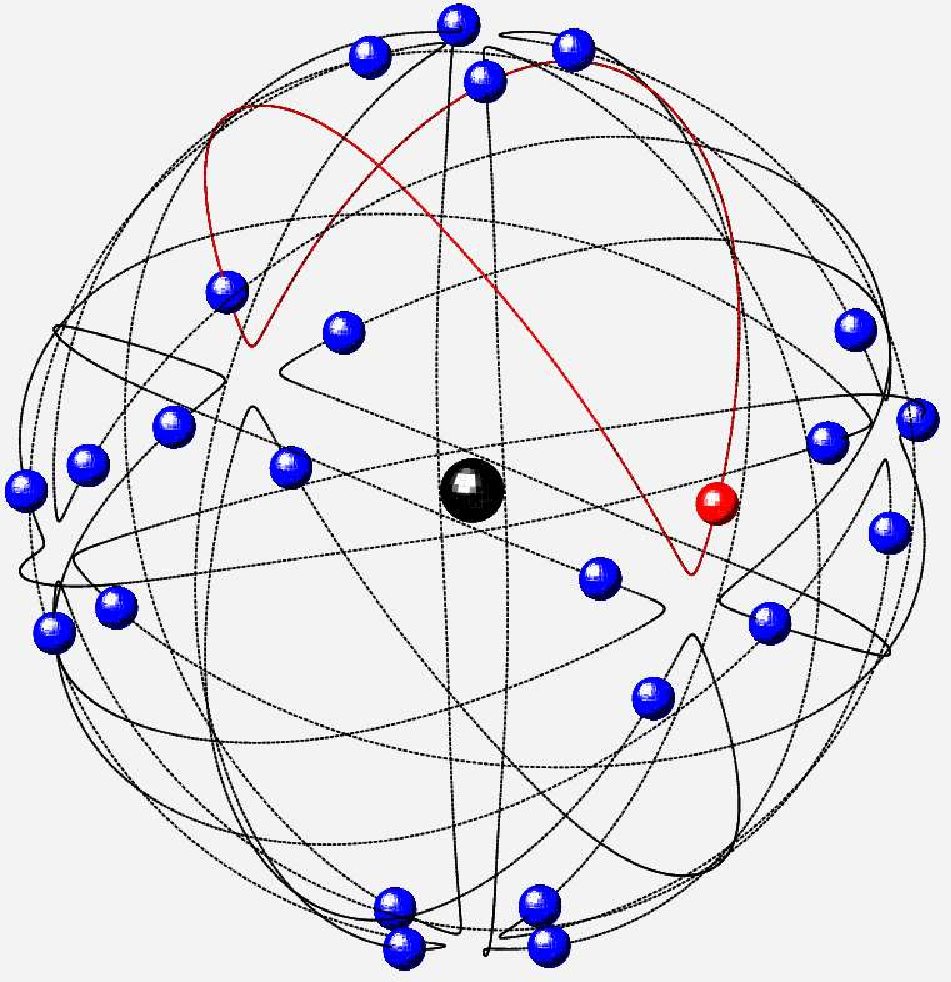}
      \hskip 0.5cm
      \includegraphics[scale=0.72]{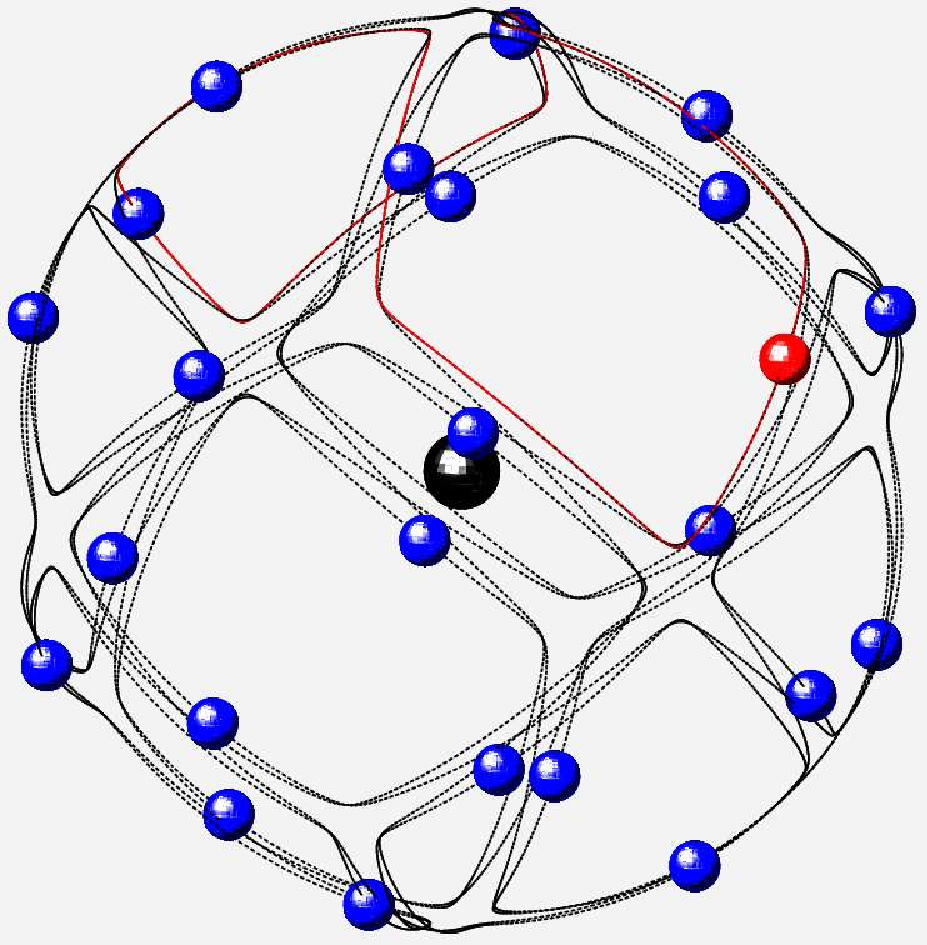}
   }
   \vskip 0.5cm
   \centerline{
      \includegraphics[scale=0.62]{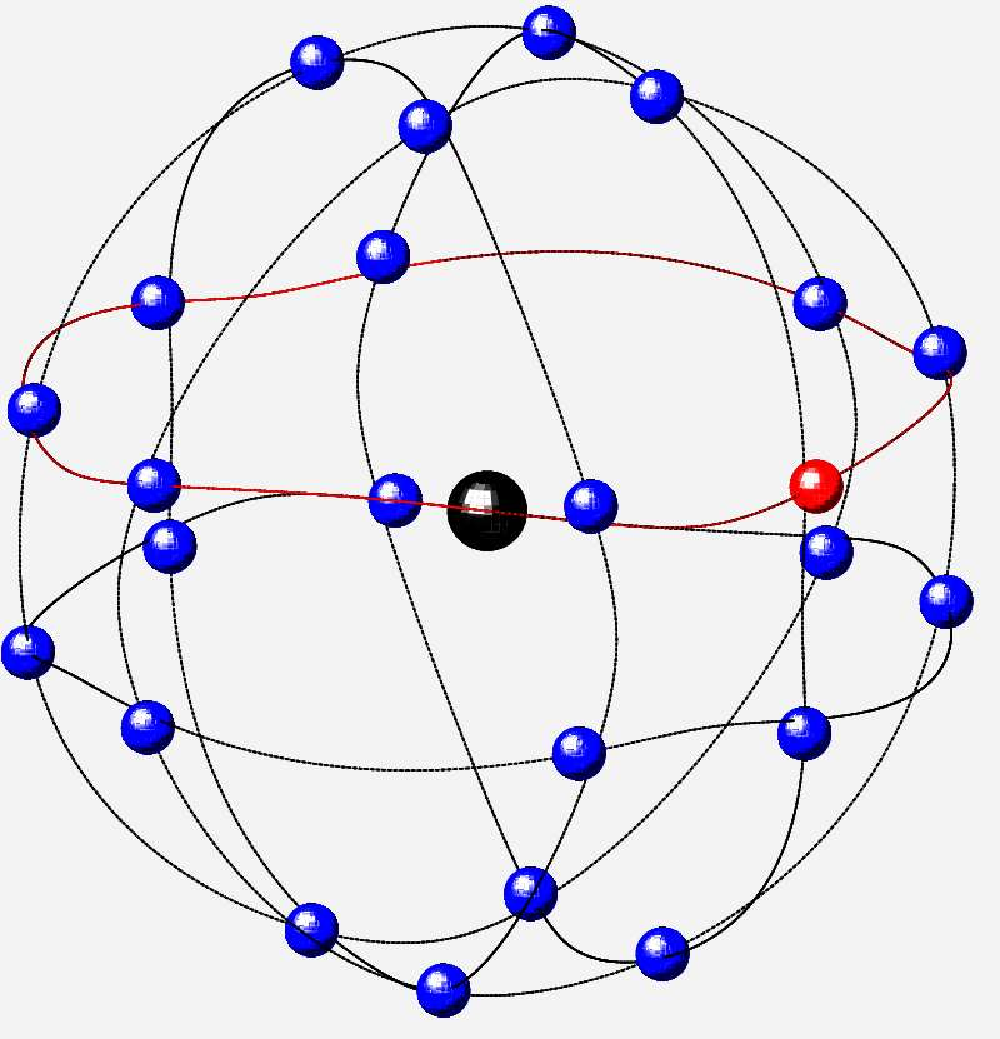}
      \hskip 0.5cm
      \includegraphics[scale=0.66]{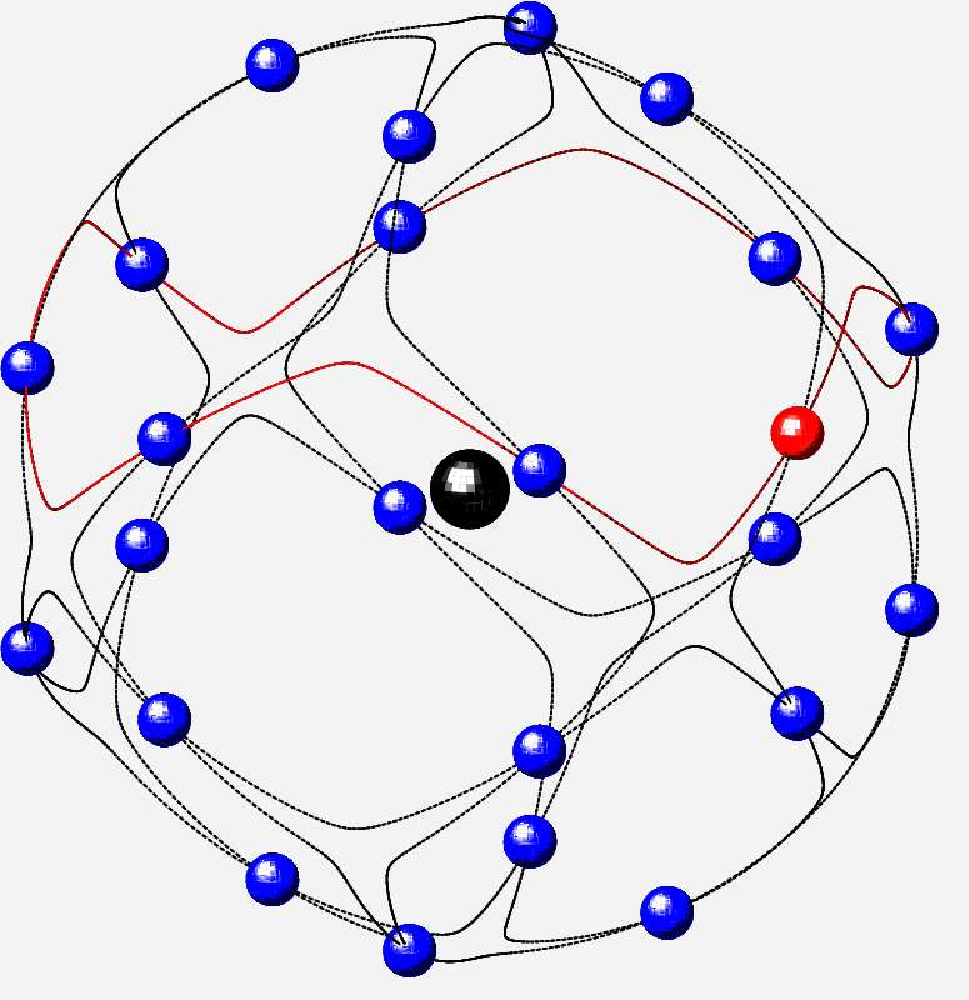}
   }
   \caption{Solutions with $24$ electrons and the symmetry of the Cube. The value of the
   central charge is $24$ for all the four examples. The homotopy class for the orbits on
   top is $\nu_{2}$ and $\nu_{43}$ for the orbits on bottom. The enumeration is referred
   to the website \cite{MFwebCoulomb}. The red electrons
represent the generating particles, and the red curve are their trajectories. The black particle
in the middle is the nucleus. The orbits on the right are obtained by the continuation
method, starting from the orbits on the left. We reached them after following a turning
point on the curve of solutions.}
   \label{fig:cubeOrbits}
\end{figure}

In \cite{FG18}, for each Platonic polyhedron, a list of free-homotopy classes 
of $\R^3 \setminus \Gamma$, each one containing a collision-free minimizer of 
the $N$-body problem with equal masses, were provided:  these lists are available at \cite{MFwebNBody}. 
Here we search for symmetric periodic solutions of the system \eqref{eq:coulombODE}, in the
same free-homotopy classes listed at \cite{MFwebNBody}. 
In \cite{FG18}, 9 and 57 homotopically different periodic orbits with the symmetry of the
Tetrahedron and the Cube, respectively, were found for the $N$-body problem with equal
masses. The total number of orbits with the symmetry of the Dodecahedron was 1442, but the
entire computation of all of them was not done.
Here we were able to compute all these orbits also in Coulomb $(N+1)$-body problem
introduced in Section \ref{s:model}, with the symmetry of the Tetrahedron and the Cube, reproducing the list in
\cite{MFwebNBody}. For the symmetry of the Dodecahedron only a few number of orbits were
computed (a large number of them is expected). Examples of orbits with
$24$ electrons are displayed in Figure \ref{fig:cubeOrbits}. More images and videos
are available at the webpage \cite{MFwebCoulomb}. 

\subsection{Continuation}
In our computations we set the period to be $T=1$: this is not
restricting, since an orbit with an arbitrary period can be found
simply by rescaling size and time.  During the continuation process we
always reached a turning point in $Q$. This means that, when we were
able to reach the physical situation of negative charged ions
(i.e. when $Q < N$), we can continue the solutions following the
turning point, and find a second orbit in which the system is neutral
(i.e. when $Q=N$).
This does not happen in all the cases we tried, and
it is not clear if there is an additional topological condition to be satisfied in order
to have the turning point below $Q=N$. 

\subsection{Stability}
\begin{figure}[!ht]
   \centering
   \includegraphics[scale=0.5]{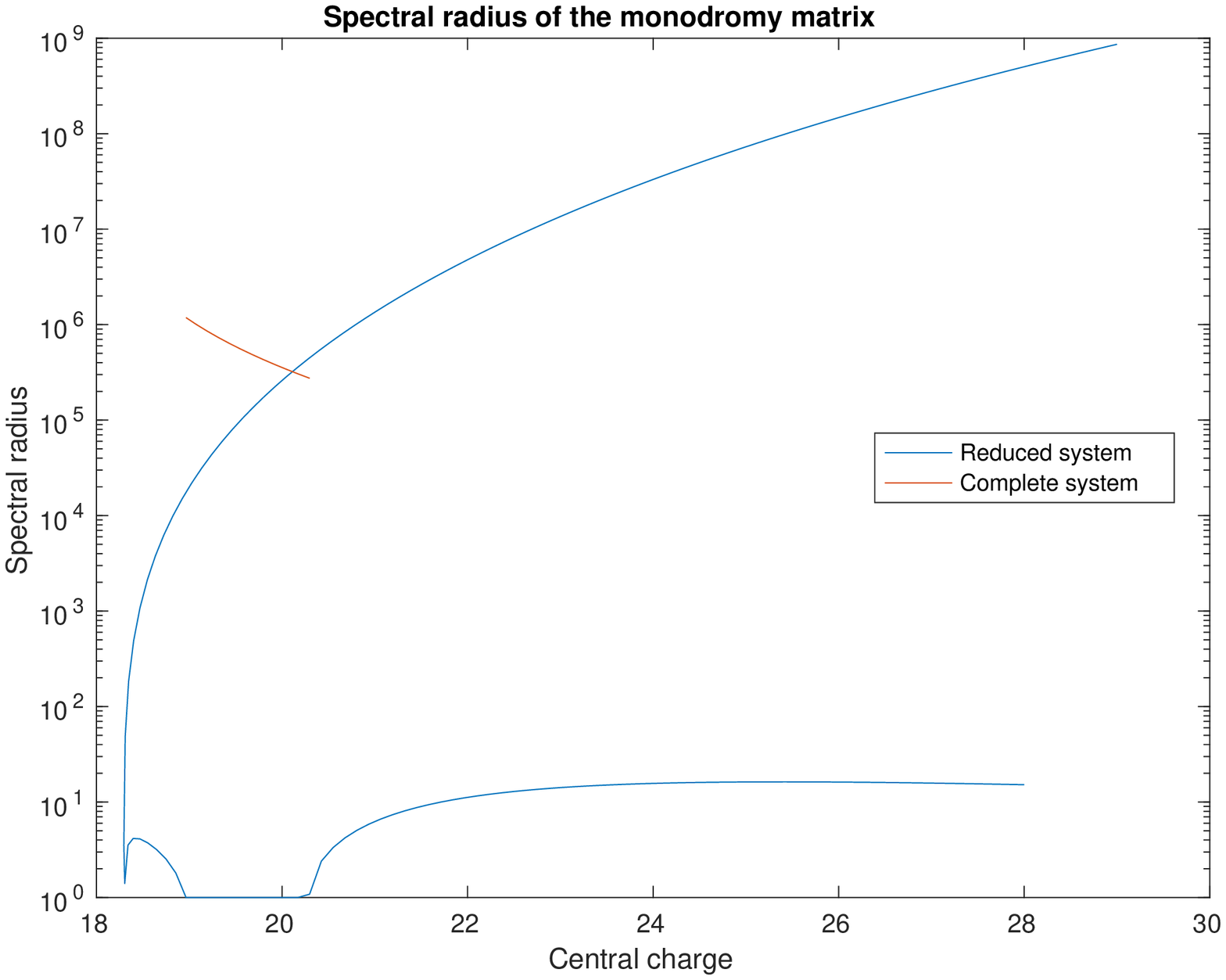}
   \caption{The spectral radius of the monodromy matrix. On the $x$ axis is reported the
   value of the central charge and on the $y$ axis is reported the value of the spectral radius, in
   logarithmic scale. The blue curve represent the evolution of the spectral radius of the
   monodromy matrix in the reduced system, while the red curve represent the evolution of the
   spectral radius in the complete system. During the continuation, the generating
   particle becomes stable for certain values of $Q$ (near $Q=20$, blue curve),
   but in fact the resulting complete orbit is unstable (red curve).  
   This plot is referred to the orbits in Figure \ref{fig:cubeOrbits}, bottom.}
   \label{fig:eigs}
\end{figure}
As said before, the study of the stability is divided in two steps: first we study the
stability of the orbit of the generating particle in the reduced system \eqref{eq:redLag},
computing a $6 \times 6$ monodromy matrix $M_6$. 
If the generating particle in unstable, then also the complete orbit in the system
\eqref{eq:coulombODE} is unstable, otherwise we proceed in the computation of the complete
$6N \times 6N$ monodromy matrix $M_{6N}$. During the continuation, the six eigenvalues of
$M_6$ move in the complex plane. For the most of the orbits, looking at the eigenvalues of 
$M_6$ was enough to conclude the instability, since during the continuation
a very large Floquet multiplier (of the order that ranges from $10^6$ to
$10^{20}$, depending on the orbit) appears. However, it can happen that for certain
values of the central charge, the eigenvalues of $M_6$ are all on the unit circle, meaning
that the generating particle is stable in the reduced system. In
these few cases we computed the matrix $M_{6N}$, verifying that the complete orbit is
indeed unstable, since a large Floquet multiplier arises. An example of this situation is 
reported in Figure \ref{fig:eigs}. More figures of this kind can be found at
\cite{MFwebCoulomb}. From the computations, it results that all the orbits are unstable.
Results for the orbits with the symmetry of the Tetrahedron are summarised in Table
\ref{seqlist}.

\begin{table}
  \begin{center}
  \begin{tabular}{c|ccccc}
     label  &  $\min Q$ & $|\lambda| \, (Q=\min Q)$ &
     $|\lambda_{6N}| \, (Q=\min Q)$    &   $|\lambda| \, (Q=12)$  \cr
    \hline
    $\nu_{1}$ & $10.346970805$ & $7.3761606$              & /                       & $0.18050567\cdot10^4$    \\
    $\nu_{2}$ & $8.2812509283$ & $2.5911113$              & /                       & $0.26940597\cdot10^4$    \\
    $\nu_{3}$ & $9.9652582174$ & $3.7293607$              & /                       & $0.49658024\cdot10^2$    \\
    $\nu_{4}$ & $12.905694682$ & $1.0$                    & $0.4930569\cdot10^{5}$  & /                        \\
    $\nu_{5}$ & $8.7205631222$ & $0.28289539\cdot 10^{3}$ & /                       & $0.33578812\cdot10^9$    \\
    $\nu_{6}$ & $9.5460053624$ & $0.49445974\cdot 10^{2}$ & /                       & $0.27073516\cdot10^5$    \\
    $\nu_{7}$ & $8.0760877496$ & $0.18090607\cdot 10^{5}$ & /                       & $0.33898930\cdot10^{13}$ \\
    $\nu_{8}$ & $12.905977225$ & $2.8508903$              & /                       & /                        \\
    $\nu_{9}$ & $12.905656346$ & $1.0$                    & $0.2461446\cdot10^{10}$ & /                        \\
    \hline
  \end{tabular}
  \end{center}
  \caption{Some numerical values obtained for the orbits with the symmetry of the
     Tetrahedron. Second column contains an approximation of the minimal value of the
     central charge $Q$ obtained during the continuation. Third column contains the
     spectral radius of the monodromy matrix $M_6$ for $Q=\min Q$. When this value is
     equal to $1$, we report the spectral radius of the complete monodromy matrix
     $M_{6N}$ in column four. Last column contains the value of the spectral radius of
     $M_6$ for $Q=12$, for which the system is neutral.
     The labels correspond to the enumeration used in the website \cite{MFwebCoulomb}. }
  \label{seqlist}
\end{table}

\subsection{Are these orbits minimizers?}
To understand better the variational nature of the orbits computed, we wonder
whether they are minimizers of the action \eqref{eq:actionSym} or not. Verify that the
orbits are global minimizers is hard to do with only numerical methods, since all
the loops have to be taken into account. However, we can verify if they are at least
directional local minimizers, weak local minimizers or strong local minimizers. To this
end, we recall here briefly the definitions and the results that we need.

\paragraph{Formulation of the problem}
Fixed $T>0$, let us consider a functional 
\begin{equation}
   \mathcal{A}(u) = \int_{0}^{T}L(t,u,\dot{u}) dt,
   \label{eq:funct}
\end{equation}
where $L: [0,T] \times \Omega \to \R$ is a $C^2$ function, $T$-periodic in the variable
$t$, and $\Omega \subseteq \R^n \times \R^n$ is an open set. 
We define the space of the $T$-periodic functions
\[
   V = \{ u \in C^1([0,T], \R^n) : u(0) = u(T) \}.
\]
and we consider $\A$ defined on a subset $X \subseteq V$.
\begin{definition}
   We say that $u_0 \in X$ is a 
   \begin{itemize}
   \item[(GM)] \textit{global minimum point} if $\A(u) \ge \A(u_0)$ for all $u \in X$;
   \item[(SLM)] \textit{strong local minimum point} if there exists $\varepsilon > 0$ such
       that for all $u \in X$ satisfying
      \[
         \norm{u-u_0}_\infty < \varepsilon;
      \]
      we have that $\A(u) \ge \A(u_0)$;
   \item[(WLM)] \textit{weak local minimum point} if there exists $\varepsilon > 0$ such
      that for all $u \in X$ satisfying
      \[
         \norm{u-u_0}_\infty + \norm{\dot{u}-\dot{u}_0}_\infty  < \varepsilon;
      \]
     we have that $\A(u) \ge \A(u_0)$; 
  \item[(DLM)]\textit{directional local minimum point} (DLM) if the function 
      \[
         \varphi(s) := \A(u_0+sv),
      \]
      has a local minimum point at $s=0$ for all $v \in V$. Note that, fixed
      $v \in V$, $\varphi:(-\delta,\delta) \to \R$ is a function of the real variable $s$.
   \end{itemize}
\end{definition}
It is clear that (GM) implies (SLM), which implies (WLM), which implies (DLM). 
Moreover, it is known that a necessary condition for a regular function $u_0$ to be a (DLM) is that it
solves the Euler-Lagrange equation associated to \eqref{eq:funct}, i.e.
\begin{equation}
   \frac{d}{dt} L_{\dot{u}}(t, u_0(t),\dot{u}_0(t)) = L_u(t, u_0(t),\dot{u}_0(t))
   \label{eq:EL}
\end{equation}
Note also that a solution $u_0$ of \eqref{eq:EL} is a (DLM) if and only if the second variation
\[
   \delta^2 \A(v) = \int_{0}^{T} \big(v(t) \cdot \hat{L}_{uu}(t) v(t) + 2 \dot{v}(t)\cdot
   \hat{L}_{u\dot{u}}(t) v(t) + \dot{v}(t)\cdot \hat{L}_{\dot{u}\dot{u}}(t)\dot{v}(t) \big) dt,
\]
is non-negative for all $v \in V$, where
\begin{gather*}
   \hat{L}_{uu}(t) = L_{uu}(t,u_0(t), \dot{u}_0(t)), \\ 
   \hat{L}_{u\dot{u}}(t) = L_{u\dot{u}}(t,u_0(t), \dot{u}_0(t)), \\ 
   \hat{L}_{\dot{u}\dot{u}}(t)= L_{\dot{u}\dot{u}}(t,u_0(t), \dot{u}_0(t)). 
\end{gather*}
The second variation is a \textit{quadratic functional}. Necessary and sufficient
conditions for a quadratic functional to be positive definite are given in
\cite{dosla:quadfunc}, for general boundary conditions. We recall briefly here the main
theorem and the definitions needed to state it.

\paragraph{Quadratic functionals}
Let $[a,b]\subseteq \R$ be a closed interval, we consider a general quadratic functional
\begin{equation}
   \Q(v) = \int_{a}^{b} \big( v\cdot P(t) v + 2 \dot{v}\cdot Q(t) v + \dot{v}\cdot R(t)
   \dot{v} \big) dt,
   \label{eq:quadfunc}
\end{equation}
where $P,Q,R:[0,T] \to \R^{n\times n}$ are $C^1$
matrix functions such that $P(t)=P^T(t), \, R(t)=R^T(t)$
for all $t \in [0,T]$.
Given a matrix $D \in \R^{2n \times 2n}$, we consider $\Q$ defined on functions $v:[a,b] \to
\R^n$ such that
\begin{equation}
   D
   \begin{pmatrix}
      v(a) \\ v(b)
   \end{pmatrix}=0.
   \label{eq:quadBC}
\end{equation}
The Euler-Lagrange equation associated to \eqref{eq:quadfunc} is 
\[
   \frac{d}{dt}[R\dot{y} + Qy] = Q^T \dot{y} + Py,
\]
and it is usually called \textit{Jacobi differential equation}.
If $\det R(t) \neq 0$ for all $t \in [0,T]$, we can write the system as
\begin{equation}
   \begin{cases}
      \dot{y} = Ay+Bz, \\
      \dot{z} = Cy-A^Tz,
   \end{cases}
   \label{eq:jde1}
\end{equation}
where 
\[
   A = -R^{-1}Q, \quad B=R^{-1}, \quad C = P-Q^TR^{-1}Q.
\]
Note that $B, C$ are symmetric matrices.
It is useful to define also the matrix version of the equation, i.e.
\begin{equation}
   \begin{cases}
      \dot{Y} = AY+BZ, \\
      \dot{Z} = CY-A^TZ,
   \end{cases}
   \label{eq:jde2}
\end{equation}
where $Y, Z: [0,T] \to \R^{n \times n}$ are matrix functions.
We introduce now some conditions and give some definitions, useful to state the main
theorem.
\begin{definition}
   Let $(y,z)$ be a solution of system \eqref{eq:jde1} such that $y(a)=0$. A point $c \in
   (a,b]$ is said to be \textit{conjugate} with $a$ if 
   \[
      y(c)=0.
   \]
\end{definition}
\begin{definition}
We say that the \textit{strengthened Legendre condition} (L') holds if $R(t) >
0$ \footnote{In the following, when we write $A>0$ ($A \geq 0$), where $A \in \R^{n \times n}$ is a symmetric
matrix, we mean that $A$ is positive definite (positive semi-definite).} for all $t \in
[a,b]$ for all $t \in [a,b]$.
\end{definition}
\begin{definition}
We say that the \textit{strengthened Jacobi condition} (J') holds if every solution $(y,z)$ of
\eqref{eq:jde1} with initial condition $y(a) = 0$ does not have any conjugate point
$c \in (a,b]$ with $a$.
\end{definition}
Note that condition (J') is equivalent in saying that the solution
$(Y_a,Z_a)$ of \eqref{eq:jde2} with initial conditions
\[
\begin{cases}
   Y_a(a) = 0, \\
   Z_a(a) = \Id,
\end{cases}
\]
is such that $\det Y_a(t) \neq 0 $ for $t \in (a,b]$. The following theorem gives
necessary and sufficient conditions for $\Q$ to be positive definite.
\begin{theorem}
   Let condition (L') hold. We have that 
   \[
      \Q(v) > 0,
   \]
   for any nonzero $v$ satisfying \eqref{eq:quadBC} if and only if condition (J') holds and 
   \begin{equation}
      \alpha^T
      \begin{pmatrix}
         -W_b(a) & -Y_{a}^{-1}(b) \\
         -Y_{a}^{T-1}(b) & W_a(b)
      \end{pmatrix}
      \alpha > 0
      \label{eq:condIII}
   \end{equation}
   for all nonzero $\alpha \in \R^{2n}$ such that $D\alpha = 0$. Here
   $(Y_a,Z_a), \, (Y_b, Z_b)$ are the solutions of \eqref{eq:jde2} given by the initial
   conditions
   \[
      \begin{cases} 
         Y_a(a) = 0,\\
         Z_a(a) = \Id,
      \end{cases} 
      \qquad 
      \begin{cases}
         Y_b(b) = 0,\\ 
         Z_b(b) = -\Id, 
      \end{cases}
   \]
   and $W_a = Z_a Y_a^{-1}, \, W_b = Z_b Y_b^{-1}$.
   \label{th:DDsufcond}
\end{theorem}

\subsubsection*{The case of symmetric orbits in the Coulomb $(N+1)$-body problem}
We take into account the symmetry of the space of loops in the theory summarized
above modifying the minimization problem.
Let $u :[0,T] \to \R^{3N}$ be a loop satisfying condition (a) and the additional
choreography constraint
\[
   u_I\bigg(t +\frac{T}{M}\bigg) = R u_I(t), \quad t \in [0,T],
\] 
for a given $R \in SO(3)$ and $M \in \N$. For the sake of simplicity, we will work with
the function $u_I$, which represents the motion of a single electron along the periodic
orbit.
From expression \eqref{eq:actionSym}, we have that
\begin{equation}
   \int_{0}^{T} L(u_I,\dot{u}_I)\,dt=M\int_{0}^{\frac{T}{M}}L(u_I,\dot{u}_I) \, dt.
   \label{eq:actionChoreo}
\end{equation}
We consider the functional
\[
   \bar{\A}(u_I) = \int_{0}^{T/M}\bigg( \frac{1}{2}|\dot{u}_I|^2 - \frac{1}{2}
   \sum_{R \in \mathcal{R} \setminus\{ I\}} \frac{1}{|(R-I)u_I|} + \frac{Q}{|u_I|} \bigg)
   \,
   dt,
\]
defined on the set of loops 
\[
   u_I:\bigg[0,\, \frac{T}{M} \bigg]\longrightarrow \R^{3},
\]
such that $Ru_I(0) = u_I(T/M)$. 
Note that, by means of \eqref{eq:actionChoreo}, if $u_I^*:[0,T] \to \R^3$ is a minimizer of the functional $\A$, then
the restriction 
\[
   u_I^*\big\lvert_{[0,T/M]}:\bigg[0,\,\frac{T}{M}\bigg]\longrightarrow \R^{3},
\]
is a minimizer of $\bar{\A}$. Vice versa, if
$u_I^*:[0,T/M]\to \R^{3}$ is a minimizer of $\bar{\A}$, then we can extend it to a
closed loop $u_I^*:[0,T] \to \R^{3}$, simply by using the rotation $R$, and we obtain a
minimizer for $\A$. 

Therefore, for the functional $\bar{\A}$, we have that $[a,b] = [0,T/M]$. Moreover, the matrix $D$ defining the
admissible curves for the second variation is
\[
    D= 
    \begin{pmatrix}
       R & -\Id \\
       0 & 0
    \end{pmatrix} \in \R^{6\times6}.
\]
Therefore, a vector $\alpha\in \R^{6}$ satisfying $D\alpha=0$ is of the form
\[
   \alpha =
   \begin{pmatrix}
      \beta \\ R\beta
   \end{pmatrix},
\]
where $\beta \in \R^3$. Inserting this relation in \eqref{eq:condIII}, we obtain that the
second variation associated to a solution of Euler-Lagrange equation $u_0$
is positive definite if and only if (J') holds and the $3\times3$ matrix
\begin{equation}
   -W_{T/M}(0) - Y_0^{-1}(T/M)R - R^T Y_0^{T-1}(T/M) + R^TW_0(T/M)R
   \label{eq:condIIIadapted}
\end{equation}
is positive definite. Note that condition (L') is always satisfied, since we have that
\[
   \frac{\partial^2 L}{\partial \dot{u}_I^2} = \Id.
\]

\begin{figure}
   \centerline{
      \includegraphics[scale=0.35]{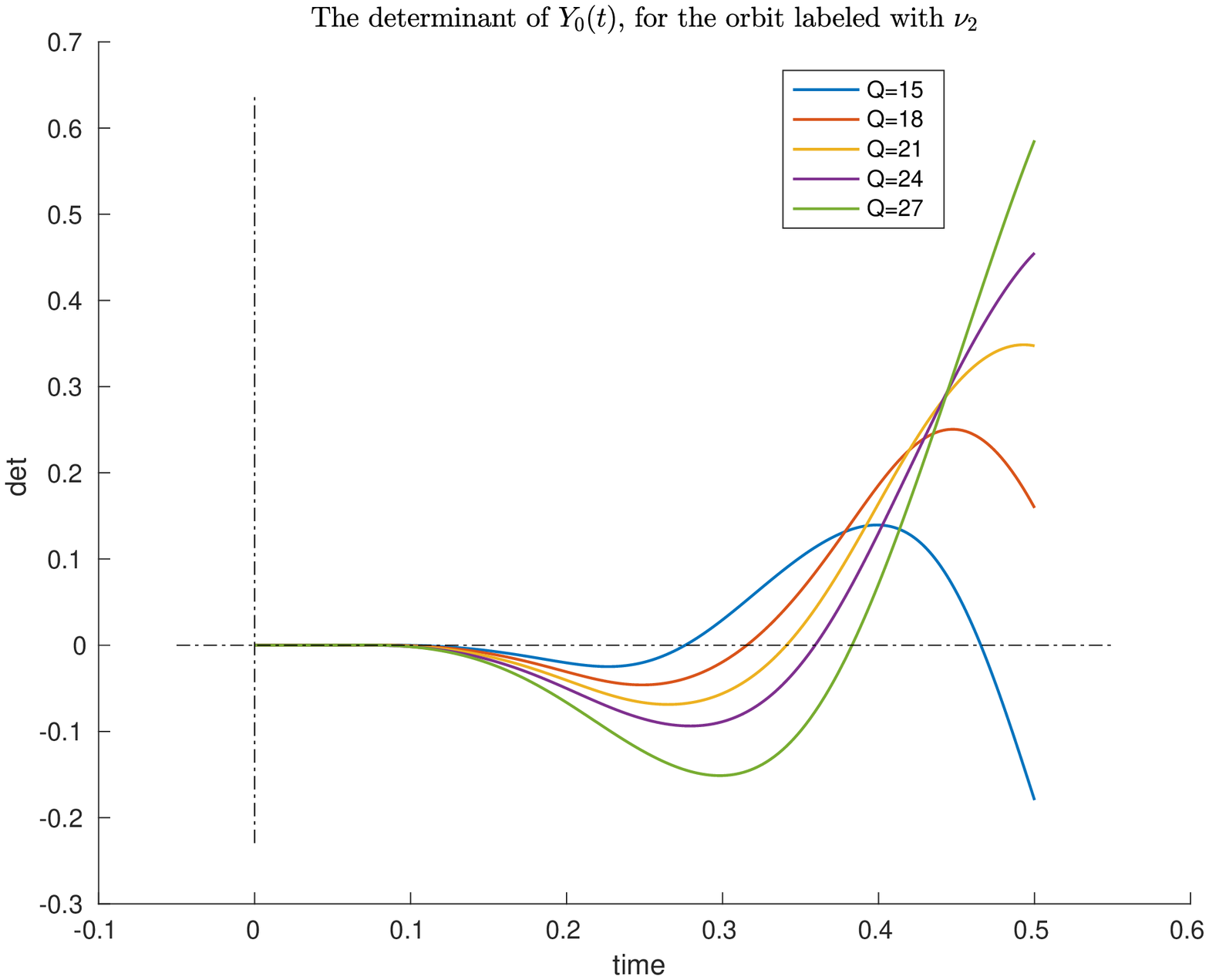}
      \hskip 0.5cm
      \includegraphics[scale=0.35]{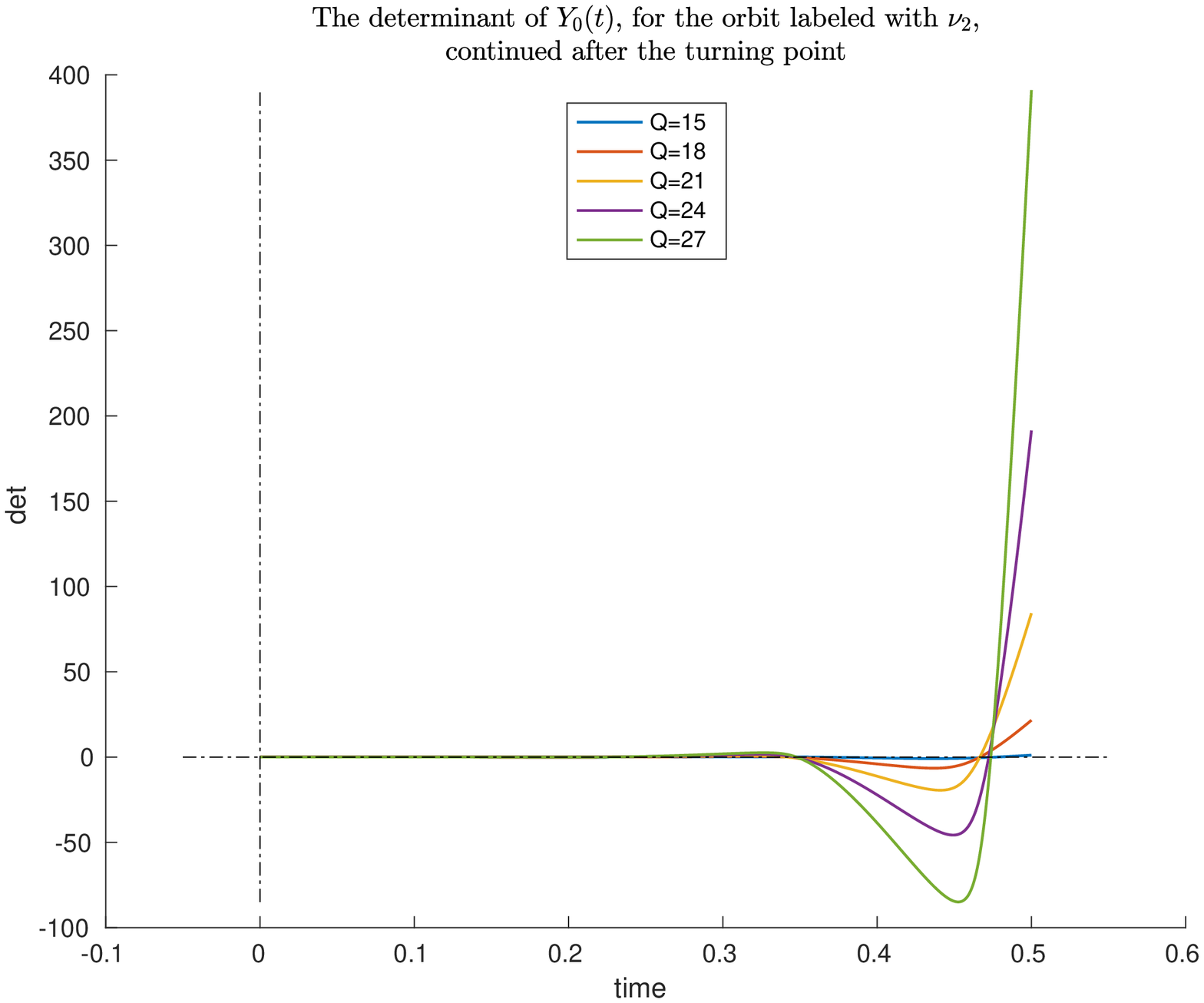}
   }
   \caption{The determinant of the matrix $Y_0(t)$ in the fundamental interval $[0,T/M]
   =[0,1/2]$, for different values of the central charge $Q$. These plots are referred to
   the periodic orbits in Figure~\ref{fig:cubeOrbits}, top.}
   \label{fig:detTrend_nu2}
\end{figure}

\begin{figure}
   \centerline{
      \includegraphics[scale=0.35]{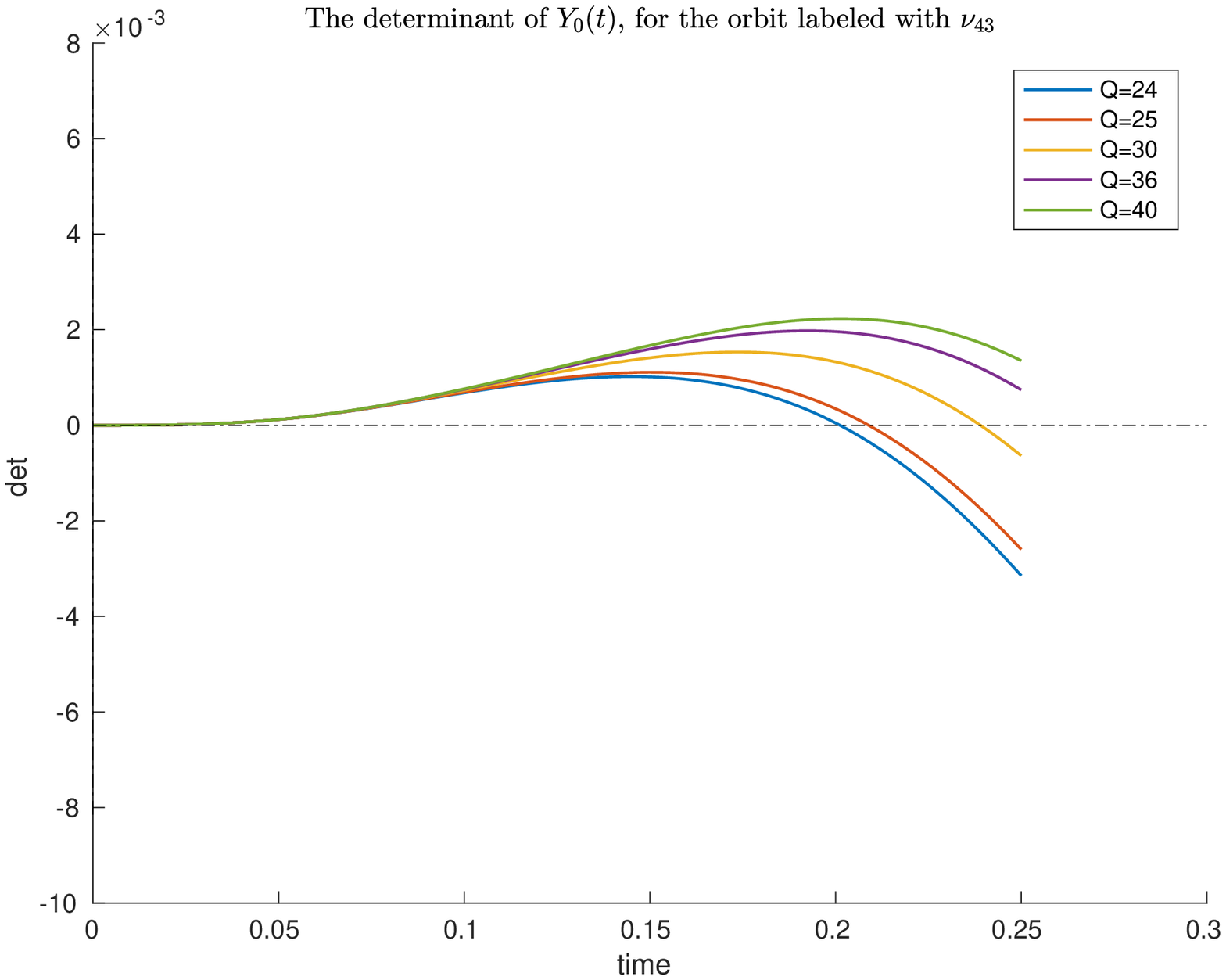}
      \hskip 0.5cm
      \includegraphics[scale=0.35]{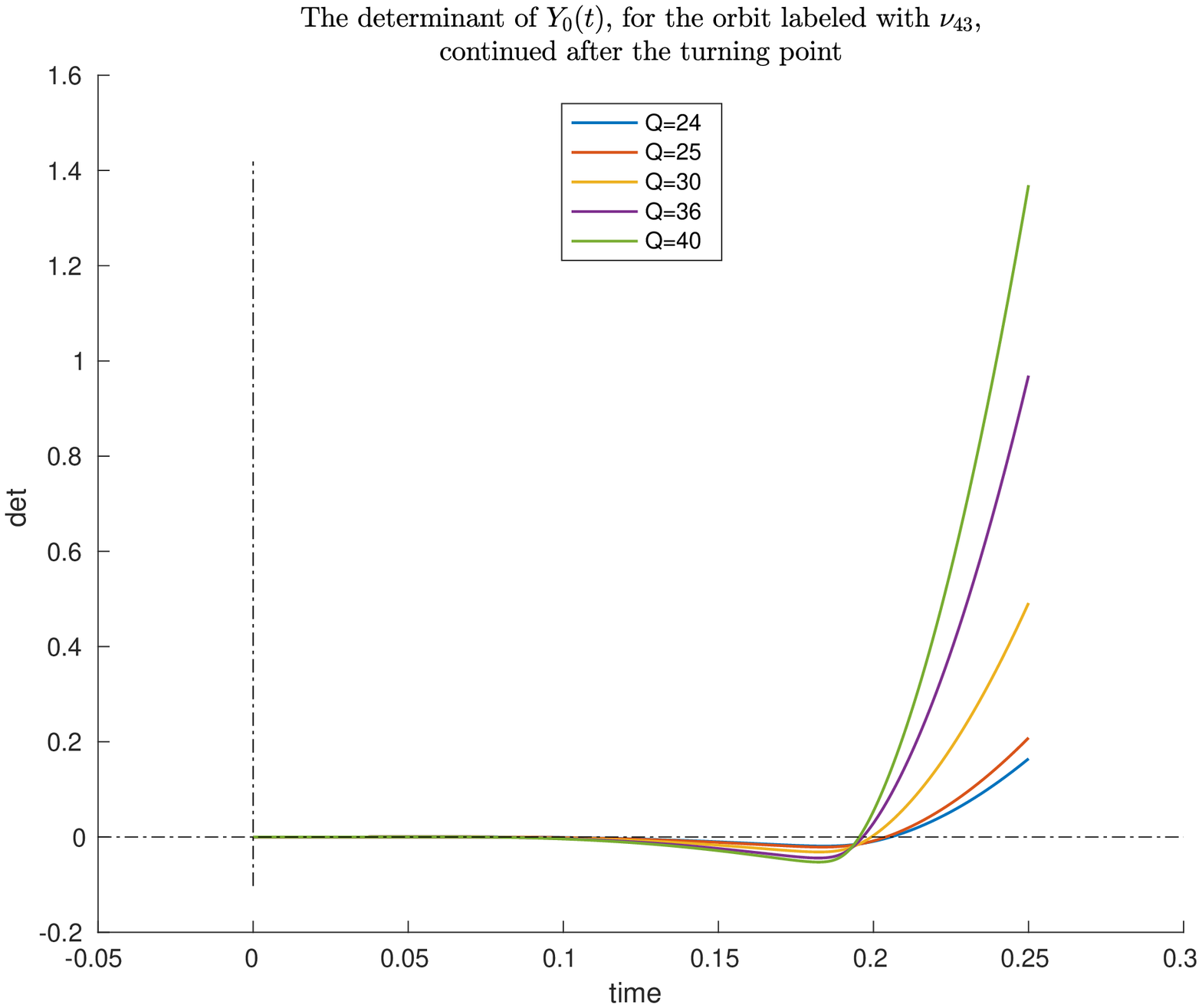}
   }
   \caption{The determinant of the matrix $Y_0(t)$ in the fundamental interval $[0,T/M]
   =[0,1/2]$, for different values of the central charge $Q$. These plots are referred to
   the periodic orbits of Figure~\ref{fig:cubeOrbits}, bottom.}
   \label{fig:detTrend_nu43}
\end{figure}

To decide whether a solution that we compute is actually a local minimizer or not, we
check if it is a (DLM) or not.
To do so, we search for conjugate points in the fundamental interval $[0,T/M]$, simply
by computing the solution $(Y_0,Z_0)$ of \eqref{eq:jde2} with initial conditions
\[
   \begin{cases}
Y_0(0)=0, \\ Z_0(0) = \Id,
   \end{cases}
\]
and then plotting the determinant of $Y_0(t)$. Since the computation is quite fast, we can also see
how the determinant evolves with respect to the value of the central charge $Q$, 
including its computation in the continuation process.  
Most of the orbits computed have a behaviour similar to
the one shown in Figure~\ref{fig:detTrend_nu2}, i.e. they have at least a conjugate point in
the fundamental interval $[0,T/M]$, indicating that they are not minimizers, not even
directional. Moreover, after the turning point the presence of a conjugate point seems to
be more likely, also because we saw that the value of the action of the periodic orbit generically increase
with respect to the previous orbit with the same value of $Q$.

However, it can occur that, during the continuation process, the determinant of
$Y_0(t)$ does not vanish for certain values of the central charge $Q$. For example, in
Figure~\ref{fig:detTrend_nu43}, we can see that the determinant is positive for the
values $Q=36,40$. Continue increasing the value of the charge $Q$, this behaviour still 
persists, and it seems that the determinant has a limiting curve that does not vanish in
the fundamental interval $(0, 1/4]$. 
Hence we also have to compute the matrix in \eqref{eq:condIIIadapted}, and verify whether it is
positive definite or not. In this case, the eigenvalues of the matrix in
\eqref{eq:condIIIadapted}, for $Q = 40$, are
computed to be
\[
    14.723038, \quad
    5.5236623, \quad
   -307.98056,
\]
hence this orbit is also not a local minimizer, despite the absence of conjugate points.
For values of $Q>40$, this property still holds, and the negative eigenvalue seems to
converge to a value close to $-60.757245$.

Further computations for the remaining orbits show that the two described behaviours are 
common to all of them, suggesting that they are not local minimizers, but indeed different
kind of stationary points, such as saddles. For this reason the method of minimization of
the action does not seem to work for the Coulomb $(N+1)$-body problem to find periodic
orbits, and maybe other variational techniques have to be used to provide a rigorous proof
of their existence.



\section*{Acknowledgments}
The first author acknowledges the project MIUR-PRIN 20178CJA2B titled ``New 
frontiers of Celestial Mechanics: theory and applications''. 
The second author has been supported by the Spanish grants 
PGC2018-100699-B-I00 (MCIU/AEI/ FEDER, UE) and the Catalan grant 2017 
SGR 1374. The project leading to this application has received 
funding from the European Union's Horizon 2020 research and innovation 
programme under the Marie Sk\l{}odowska-Curie grant agreement No 
734557. 

\bibliography{mybib}{} 
\bibliographystyle{plain}
\end{document}